# Characteristic Study and Development of Surface Resistivity Measuring Device for Resistive Plate Chamber Detector

A. Kumar[1], A. Pandey[1], N. Marimuthu[1,2], M. K. Singh[1], , V. Singh[*1] , and S. S. R. Inbanathan[2]

[1]*Department of Physics, Institute of Science, Banaras Hindu University, Varanasi - 221005, India*
[2]*Department of Physics, Post Graduate and Research, The American college Madurai-625002, India*
*E-mail*: `venkaz@yahoo.com`

ABSTRACT: The **I**ndia-based **N**eutrino **O**bservatory (INO) is an approved multi-institutional experiment of India. It will use maximum 30,000 **R**esistive **P**late **C**hamber (RPC) of size 2×2m as an active detector in which each will be sandwiched between two iron plates. Each RPC detectors will be made up of 3 mm thick glass (or Bakelite) plates whose one surface will be coated with graphite paint to work as electrode. To achieve effective and uniform electric field, the coated graphite paint must have uniform distribution over the glass surface. Uniformity can be demonstrated by measuring the surface resistance of the coated graphite paint. However, manual measurement of 60,000 electrodes with the help of multimeter and the jig demands huge amount of person-hour with compromised accuracy, and therefore that requires the need and importance of the automatic scanning system. An advanced automatic scanning system (ASS) has been designed & developed to fulfill the required needs and overcome the shortcomings of the initially developed system. Measurement are sensitive to the contact pressure between the jig and RPC electrode's surface. In this paper, various parameters of the ASS have been precisely fixed. Obtained results are compared with the standard one and found satisfactory within error. It has been observed that for obtaining the correct and safe value of surface resistance, applied pressure value in terms of force should be in the range of 9-11 Newton.

# Contents

## 1. Introduction

The **I**ndia-based **N**eutrino **O**bservatory (**INO**) is one of the approved mega science project of India. It will use nearly 30,000 RPC detectors during full operation, which will be the world's first experiment to use largest number of RPC detectors. Each RPC detectors consist of two glass (or Bakelite) electrodes. RPC electrodes are developed using graphite paint coating over a surface of the glass (or Bakelite). Graphite paint coating must be as much uniform as possible, because it will affect the uniformity of the applied electric field on the chamber. Therefore, uniformity of coated graphite paint is one of the most important criteria in the fabrication of quality RPC detector, as shown in figure 1 (b). Surface resistivity measurement of the electrodes is manually performed in which a square shape jig is placed over the surface of electrode and records resistance value using multimeter [1]. However, manual measurement of the surface resistance of 60,000 electrodes of size 2×2m is time consuming, very challenging and creates lots of issues. In manual measurement, the key factor is the correctness of the observed value because that depends upon the applied pressure on the jig during measurement. However, applied pressure is not constant in manual measurements which introduce significant error in the observed value. The jig is an arrangement designed and developed indigenously consisting of a square sized insulating plate having copper strips adhered on the opposite edges of the plate over the soft foam [2-4]. During measurement, the placement of the jig is over the electrode's surface and should be in well contact with the surface and optimized.

This research article is based on the design and development of an instrument that can solve the afore-mentioned problems. In this connection, a prototype semiautomatic system of size 50× 50cm was designed and developed [1]. Final version of advanced automatic scanning system as shown in Fig. 1(a) is designed and developed according to the requirements of the INO collaboration of size of 2.25×2.25m. The developed advanced scanning system is cost effective and lightweight because it consists of Aluminium bars, hard rubber belt and Arduino Microcontroller along with Arduino motor shield that is easily available in local market to control the movement of all the DC motors involved.

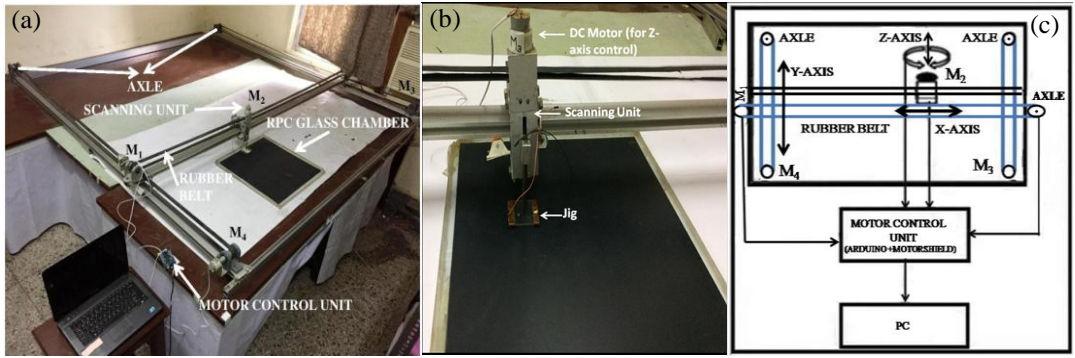

**Fig. 1:** Photographs of developed full scale automatic scanning system during data taking (a) and zoomed photo of scanning unit is displayed (b), and (c) a schematic diagram of the design [1].

Fig. 1(b) shows the zoomed portion of the scanning unit along with an upper surface of the graphite-coated glass electrode. Extension of the prototype system emerges as full scale advanced automatic scanning system, which is comfortable with the size of 2×2m electrodes. The presented results in this paper are based on the measurement with the size of 50×50cm electrodes otherwise specifically mentioned.

## 2. Design and development of automatic scanning system

### 2.1 Designing of scanning System

Automatic Scanning System (ASS) is developed using simple Aluminum bars, 12V DC (Direct Current) motors, rolling axle, rubber belt, jig and a pressure sensor. Fig. 1(c) shows schematic diagram of the design. In the design, the moving arms control the surface scanning of the electrodes along the x- and y-axis with the help of DC motors namely $M_1$, $M_3$ and $M_4$. An active component of the developed ASS is the scanning unit called "Jig" and used for the measurement of surface resistivity. Jig also works with the help of two DC motors in which one motor ($M_1$) is used for its movement along the surface and another one ($M_2$) is used for the motion along the z-axis i.e. for the up and down. Jig consists of two copper strips of length 5cm and 5cm apart with insulating material and adhered on a soft-spongy material. The dimension of the jig is variable and varies according to the size of the electrodes. Complete mechanism of the automatic system is explained in Ref. [2-4].

### 2.2 Force Sensing Sensor

The surface resistivity measurements of the electrodes were performed by the ASS with the ordinary jig. The pressure applied on the jig seems to have variation and the observed values have no proof of correctness. Therefore, a Force Sensing Resistor (FSR), which will sense the contact pressure between surface of the electrode and jig, is attached in between the two strips of the jig of size 5×5cm, as shown in Fig. 2(a). The FSR consists of a piezoresistivity conductive polymer, which changes resistance by the application of force on it.

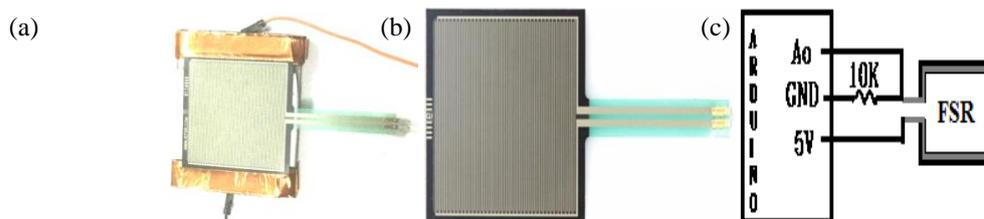

**Fig. 2:** (a) Photograph of FSR sensor affixed with the jig, (b) a square shape FSR sensor and (c) FSR activating circuit diagram [9].

The sensing film with sensing area of 2cm$^2$ consists of both electrically conducting and non-conducting particles suspended in matrix as shown in Fig. 2(b) is mounted in the empty space between the strips of jig. The FSR has been activated with the help of Arduino microcontroller (ATmega 328P) in assistance of the activating circuit as shown in the Fig. 2(c). The raw data is obtained from the FSR through analog terminal ($A_0$) of the Arduino. Coding is done in order to convert the obtained raw data into the force unit (i.e. Newton, N). The FSR sensor can sense the applied force in the range of 1 - 100N in the integral units.

## 3. Measurements, results and discussion

### 3.1 Mechanism for surface resistance measurement

The surface resistance measurement of the RPC electrode is based on the fundamental principle of voltage divider for which Arduino is used [5-9]. A computer programming code has been developed for measuring the voltage drops across the known resistance. Consequently, since voltage drop across jig is known therefore, resistance across it can easily be measured through the Arduino [5-9]. Using FSR, again a new programming code has been developed in such a way that the resistance-measuring circuit is activated only when certain force is applied on the jig. Therefore, the main advantage of the FSR is to insure that the measured resistance value is obtained when the jig touches the electrode surface at a certain calibrated pressure.

### 3.2 Calibration of FSR unit

To know the precision of the ASS in case of resistance measurement system measures the resistance value of the known resistances. For this purpose, three known resistance values such as 10kΩ, 120kΩ and 285kΩ are used. The ASS measured the value of the resistances using scanning jig at different contact force acting between jig and the electrode's surface. The obtained results have been shown in Fig. 3(a). The statistical error bar is mentioned. It can see from the plot that, the measured values of the known resistance are on average unaltered within the error bars with the variations in the force acting between jig and the known resistance terminals.

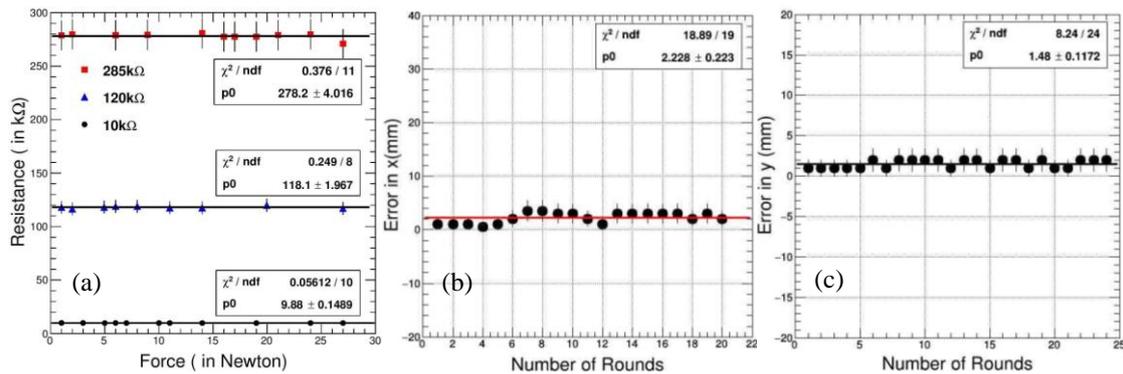

**Fig. 3:** Variation of observed resistance value with respect to the known value using FSR equipped ASS (a), observed error in the position measurement along the x– (b) and y– directions (c).

Indeed, there is a very small variation observed in the resistance under the application of heavy force, which may be the case of abnormal conditions. It means there is no effect of applied force on the value of known resistance within the workable force region.

– 4 –

## 3.3 Error in the measurement of the ASS

### 3.3.1 Position measurement

In the surface resistance measurement of glass electrode, it is necessary that the developed ASS should cover all the areas without leaving any gap. For this purpose, an experiment performed in which jig moves number of rounds on their defined path along the x- or y-axis and deviation from the path in each round is measured. Here the jig size (5cm) is the step size of moving scanner and in each round jig is travelling 4m distance. Keeping step size constant at the same track, jig's movement is repeated. After each repetition, the final position of the jig is measured from the reference point. For the scanning of 2×2m size electrode's surface a jig of 25cm size will be used in case of INO experiment. From Figs. 3(b & c), it may be seen that jig's movement along the x – axis has deviation in between 1 to 4 mm from its original position after taking twenty (20) rounds i.e. after travelling 80m distances. It means jig movement has average error rate 0.05mm per meter, which is negligible in comparison with $4m^2$ areas. The deviation observed along y– axis is better than the deviation in the x– axis as shown in Fig. 3(c). The error mention in Figs. 3(b & c) is purely statistical. This much deviation is possible due to some flexibility in the rubber belt and friction in the ball bearings of the motor's wheels.

### 3.3.2 Surface Resistance measurement of the electrode

Surface resistance measurement has been performed on the surface of 50×50cm RPC detector's electrodes using ASS. In the measurement, jig of dimension 5×5cm is used. Figures 4 show the variation of resistances throughout the surface of the electrode without (Fig. 4(a)) and with (Fig. 4(b)) the implementation of FSR. Figure 4(b) represents the resistance distribution with the application of 9-11N force. It observed that there is no visible change in the distribution above 11N force. It can also be observed from Figs. 4(a & b) that the resistance values are spreading effectively throughout the RPC electrode's surface in the range of (250-350) kΩ and (275- 325) kΩ, for without and with FSR application, respectively.

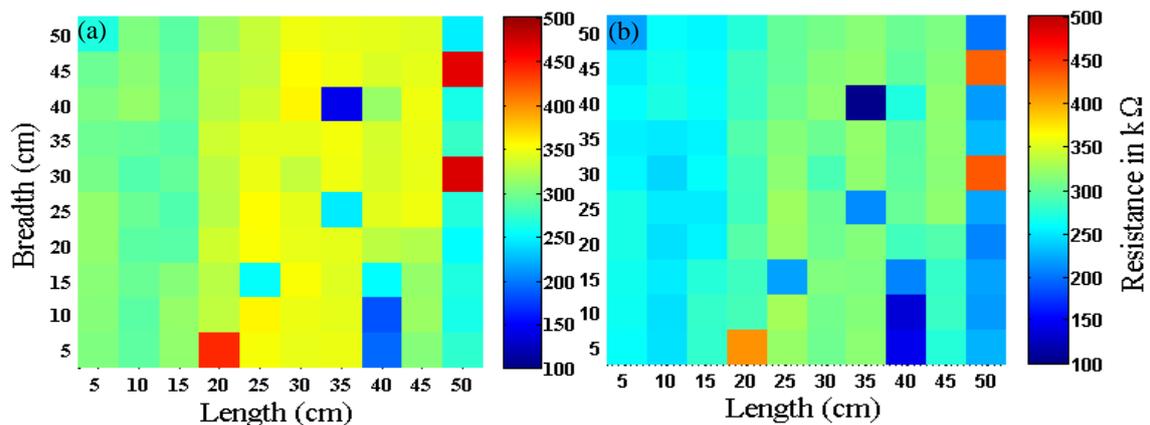

**Fig. 4:** Plots for RPC detector electrode's surface (a) without FSR and (b) with FSR having force (9-11) Newton.

It can be seen from Figs. 4 (a & b) that the area of extreme surface resistance are invariable and are mostly located around the edges of the electrode that represent a kind of validation of the measurement. Variation observed in surface resistivity towards more realistic value. In the measurement, jig is allowed to take resistance readings without contact force acting between jig and the RPC electrode's surface.

Further, it has been observed that the surface resistivity of the electrode's surface decreases as the contact force between the jig and the surface increases. Contact area between the jig and the electrode's surface increases with the force, which is imparted by the resistance-measuring jig. Therefore, more contact between jig and electrode's surface will allow a large current to flow through the electrode's surface, which results in the decrement of the surface resistance of the electrode. As the applied force ranges from (9-11N) to (11-13) Newton, no appreciable changes are observed. Therefore, it can be concluded that the contact between the jig and the surface electrode's becomes appropriate in the range of 9-11N force.

## 4. Conclusion

Advanced automatic scanning system is fully operational. The optimal value of the applied force on the glass surface to obtained more appropriate results is in the range of 9-11N. With the application of optimum pressure the observed value of the surface resistance is in the range of (275-325) kΩ within 4KΩ error. Due to the application of z-axis controller smooth movement of 0.5mm pitch is achieved, which is equivalent to a unit Newton force. This helps in performing the measurement without making any scratch on the coated paint. Therefore, we may conclude that the developed ASS is effective in terms of cost, performance and quality.

## Acknowledgments

Authors are grateful to the Department of Science and Technology (DST), New Delhi, India for providing financial support and India-based Neutrino Observatory (INO) Collaboration for their continuous support and fruitful discussions.